\documentclass[11pt,twoside]{article}
\usepackage{graphicx,epsfig,natbib,epstopdf}
\usepackage{CS18}
\begin{document}
\def\Alfven{{\rm Alfv\'en}\ }
\newcommand{\be}{\begin{equation}}
\newcommand{\ee}{\end{equation}}
\newcommand{\mb}{\mathbf}
%
%
\title{Toward A Self Consistent MHD Model of Chromospheres and Winds From Late Type Evolved Stars}
%
%
\author{V. S. Airapetian$^{1}$, J. E. Leake$^{2}$, K. G. Carpenter$^{1}$}
\affil{$^1$NASA/Goddard Space Flight Center, Greenbelt, MD 20771}
\affil{$^2$George Mason University, Fairfax, VA }
%
\begin{abstract}
We present the first magnetohydrodynamic model of the stellar chromospheric heating and acceleration of the outer atmospheres of cool evolved stars, using $\alpha$ Tau as a case study. We used a 1.5D MHD code with a generalized Ohm's law that accounts for the effects of partial ionization in the stellar atmosphere to study \Alfven wave dissipation and wave reflection. We have demonstrated that due to inclusion of the effects of ion-neutral collisions in magnetized weakly ionized chromospheric plasma on resistivity and the appropriate grid resolution, the numerical resistivity becomes 1-2 orders of magnitude smaller than the physical resistivity. The motions introduced by non-linear transverse \Alfven waves can explain non-thermally broadened and non-Gaussian profiles of optically thin UV lines forming in the stellar chromosphere of $\alpha$ Tau and other late-type giant and supergiant stars. The calculated heating rates in the stellar chromosphere due to resistive (Joule) dissipation of electric currents, induced by upward propagating non-linear \Alfven waves, are consistent with observational constraints on the net radiative losses in UV lines and the continuum from $\alpha$ Tau. At the top of the chromosphere, \Alfven waves experience significant reflection, producing downward propagating transverse waves that interact with upward propagating waves and produce velocity shear in the chromosphere. Our simulations also suggest that momentum deposition by non-linear \Alfven waves becomes significant in the outer chromosphere at 1 stellar radius from the photosphere. The calculated terminal velocity and the mass loss rate are consistent with the observationally derived wind properties in $\alpha$ Tau.
\end{abstract}

\vspace{.2in}
\normalsize

\section{Introduction}
Stars with spectral classes later than F5 (including the Sun) possess convective zones, magnetic surface activity, chromospheres and coronae. The convective zones provide the major power  source for  the UV, X-ray and radio emissions from the stellar atmospheres and are deeply connected to the initiation of the mass outflows known as stellar winds. The stellar chromosphere and transition region represent the interface layers between the photosphere and corona, and, play a critical role in specifying the amount of mechanical energy dissipating into the atmospheric heating and depositing momentum to drive stellar winds. Therefore, the stellar chromosphere regulates the mass and energy flux from the entire atmosphere and determines the dynamics and magnetic topology of the overlying layers containing the stellar wind. The net radiative flux from the chromosphere is over 10-30 times greater than that from the entire overlying corona. The problem of identifying and understanding the mechanisms which heat the outer atmospheric layers essentially is thus the same as solving the problem of chromospheric heating in stars throughout the H-R diagram.

Similar to the solar chromosphere, the chromosphere of a cool star represents a highly complex, weakly ionized and magnetized region of the atmosphere. Over two decades of observational and theoretical studies suggest that the chromospheric heating can be explained by two types of physical mechanisms: acoustic and magnetic heating (Narain \& Ulmschneider 1996). Acoustic wave energy generated by stellar convection can successfully explain "basal" flux levels in Ca II and Mg II UV emission lines (Buchholz et al. 1998). However, they are found to be incapable of explaining the magnitude of chromospheric turbulence (Judge \& Cuntz 1993; Airapetian et al. 2000) as deduced from HST/GHRS data (Carpenter et al. 1991), accelerating stellar winds, or accounting for their mass loss rates (Hartmann \& MacGregor 1980; Sutmann \& Cuntz 1995). Stellar pulsations (global oscillation modes) observed in giants can be important in forming dusty outflows in M-type Red Giant Branch stars (RGB), but seem to provide only a small contribution to the heating and momentum of atmospheres of RGB giants (Sutmann \& Cuntz 1995; Bladh et al. 2012). In contrast, magnetic energy dissipation may explain a wide range of the observed UV emission above the basal flux as well as supersonic turbulence and mass loss rates in cool giants (Suzuki 2007; Airapetian et al. 2000; 2010; Cranmer 2008; 2009; Cranmer \& Saar 2011). The presence of convection with surface magnetic fields that have been measured in a number of non-coronal and coronal giants provides an efficient source for conversion of the kinetic energy of convection into the electrical energy driving electric currents in a weakly ionized stellar atmosphere (Konstantinova-Antova et al. 2010). Such currents can be efficiently generated by MHD waves and dissipated by the resistive load of the stellar chromospheres. Specifically, the effect of ion-neutral collisions on MHD wave dissipation and the associated atmospheric heating in the atmospheres of cool evolved stars must be included. Recent models of MHD wave dissipation in the solar atmosphere suggest that the effects of ambipolar diffusion play a dominant role  (Goodman 2000; De Pontieu et al. 2001; Khodachenko et al. 2004; Leake et al. 2005; Soler et al. 2013; Airapetian \& Cuntz 2014; Airapetian et al. 2014).

In this paper, we present the first non-linear fully compressible time-dependent visco-resistive MHD model of atmospheric heating driven by \Alfven waves launched from a weakly ionized and magnetized photosphere of a giant star. Specifically, our goal is to reproduce physical conditions in the atmosphere of a K5 III giant, $\alpha$ Tau. We use the LaRe2D code with the generalized Ohm's law applied in 1.5D mode. The equations are solved for a single fluid with a generalized Ohm's law that includes ion-neutral collisions to calculate the heating rates and momentum deposition due to \Alfven waves propagating in the partially ionized stellar chromosphere described by a semi-empirical model of McMurry (1999). We restrict our study to 1.5D MHD modeling because it allows us to apply high spatial resolution to study \Alfven wave dynamics with realistic transport coefficients with fully resolved resistivity. This approach provides a realistic estimate for the heating rates of the stellar chromosphere of a red giant due to Joule wave heating. We find that non-linear \Alfven waves drive non-linear compressible MHD waves throughout the stellar chromosphere that contribute to the dissipation and acceleration of the stellar wind. We also estimate the efficiency of \Alfven wave reflection and associated ponderomotive force exerted by \Alfven wave pressure on outer atmospheric plasma and calculate associated mass loss rates from a K5 III star, $\alpha$ Tau.

\section{Observational Constrains on Stellar Chromospheric Heating}

The basic properties of stellar atmospheric heating and its dynamics in terms of outflows (terminal velocity and mass loss rate), are highly dependent on the properties and evolutionary status of the star. On the main sequence, for example, the mass loss rate can range from as high as 10$^{-6}~M_{\odot}$~yr$^{-1}$ for hot, early type stars down to 10$^{-12}~M_{\odot}$~yr$^{-1}$ or less in the cool dwarfs. For luminous evolved stars the situation becomes more complex.
The chromospheres of cool evolved stars present a case of a highly extended and supersonically turbulent  medium (2-4 times greater than the sound speed) that is signified by "quiescent" non-thermal broadening observed in SI, CII], Si II, Fe II, Co II optically thin UV emission lines, forming in an extended and rarefied stellar chromosphere (Robinson et al. 1998). The fluxes in these UV lines in some of the observed cool giants (such as $\alpha$ Tau and $\gamma$ Dra) vary with time on a scale of at least 1-2 years by a factor of 20-45$\%$ (Carpenter, Airapetian \& Kober 2013). Moreover, wind-reversed chromospheric lines, such as Mg II, show persistent blueshifts signifying mass outflow with velocities increasing with height and reaching terminal velocity of 30-70 km/s (which is greater than the stellar escape velocity) within 1 - 2$R_{\star}$ (Carpenter et al. 1995; Robinson et al. 1998). Recent observations reveal surface magnetic field at the level of a few Gauss in non-coronal giants to about 60 G in coronal giants and supergiants (Auriere et al. 2010; Konstantinova-Antova et al. 2008; 2009; 2010; 2012; Tsvetkova et al. 2013). For example, the coronal giant $\beta$ Cet shows an average bipolar photospheric field, \emph{fB}=20 G (\emph{f} is the filling factor), while coronal observations of Fe XXI lines imply magnetic confinement with coronal field of about 300 G suggesting that $\emph{f}$ is less than a few percent. Moreover, observations of UV line emission from late-type stars also strongly support the magnetic nature of stellar winds (Carpenter \& Airapetian 2009). Direct infrared VLTI/AMBER imaging  of an M type giant star, BK Virginis, has recently revealed anisotropic structures in inner regions nearer to the star which may imply wind formation in regions with open magnetic fields similar to the anisotropic solar wind forming in coronal holes (Ohnaka et al. 2013). Magnetic field should thus be seen as a critical factor in heating and depositing momentum in late type evolved stars.

In the absence of significant flows, the dissipation of chromospheric energy
due to non-radiative energy source(s) is mostly balanced by radiative cooling.
The observed surface fluxes of the two major contributors, i.e., the Mg~{\sc
ii} and Ca~{\sc ii} emission lines, allow one to define the range of required
heating rates.  Those have been given as (1-100) $\times$
10$^5$~ergs~cm$^{-2}$~s$^{-1}$ (Linsky \& Ayres 1978; Strassmeier et al.~1994;
P{\'e}rez Mart{\'{\i}}nez et al.~2011).

One-dimensional semi-empirical models of evolved stars represent powerful tools
for constraining the radial profiles of the heating rates that are related to
the deposition of energy throughout the atmosphere.  This class of model was
inspired by time-independent 1-D semi-empirical models of the solar
chromosphere developed by Vernazza et al.~(1976, 1981) and Fontenla et
al.~(2002); they were designed to reproduce the temporally and spatially
averaged UV line profiles and fluxes.  Semi-empirical models provide a
quantitative characterization of the radial profiles of temperature, electron
density, neutral hydrogen density and turbulent velocity across the atmospheres
of evolved stars.  This type of model was developed for a number of evolved
stars, such as giants like $\alpha$~Boo, $\alpha$~Tau, and $\beta$ Cet, and for
various supergiants, including the eclipsing supergiant 31~Cyg (Eriksson et al.~1983;
McMurry~1999; Eaton 2008).  A chromospheric
model for $\alpha$ Tau developed by McMurry (1999) suggests that the
temperature rises throughout the chromosphere up to 100,000\,K at about
0.2~$R_{\star}$.  At the same time, the chromosphere transitions into a wind
within one stellar radius, suggesting that the atmosphere therefore undergoes
acceleration between 0.2 and 1~$R_{\star}$.  {\it FUSE} observations of various
non-coronal giants show the presence of C~{\sc iii} and O~{\sc vi} lines,
indicating hot plasma with temperatures up to 300,000\,K.  Plasma at such high
temperatures occupies low volumes and appears to be mostly at rest with respect
to the photosphere in stars that have winds of low escape velocities,
indicating that the plasma should be magnetically confined (Ayres et al.~2003;
Harper et al.~2005; Carpenter \& Airapetian 2009).

Recent detections of surface magnetic fields for some G--M giants and
supergiants suggest that surface magnetic fields could be an important
contributor to the thermodynamics of the outer chromosphere (Auri\'ere et
al.~2010; Konstantinova-Antova et al.~2010, 2012).  The observed field
strengths vary from 0.5 to 1.5~G in late-type giants and increase to 100~G in
early-type coronal giants.  Rosner et al.~(1995) suggested that as stars evolve
toward the giant phase, their magnetic topology transitions from closed
magnetic configurations to predominantly open ones; the latter allow massive,
non-coronal winds to be supported.  If the magnetic field is non-uniformly
distributed over the stellar surface, the associated radial profiles in the
atmosphere can be determined by assuming that the magnetic pressure inside an
untwisted (purely longitudinal) flux tube, ${B^2}/{8\pi}$, is balanced by the
gas pressure of the surrounding non-magnetic atmosphere, $P_{\rm ext}$. This
suggests that the plasma pressure inside the tube is smaller than the magnetic
pressure of the plasma, $\beta = {7 n_9 T_4}/{B_1^2}$, where
$n_9=n/10^9$~cm$^{-3}$, $T_4=T/10,000$\,K, and $B_1=B/10$~G.  For typical
chromospheric conditions of $n_9 \sim$ 1 and $T_4\sim$1,
the plasma-$\beta$ becomes less than 1 at B $\geq$ 50 G.  Observations in the
vicinity of active regions on the Sun that are represented by plages indicate
magnetic fields of a few hundred Gauss at chromospheric densities and
temperatures; the force balance between the magnetic and plasma pressures can
therefore be described satisfactorily by the thin flux-tube approximation
(Rabin 1992; Gary 2001; Steiner 2007; Judge et al. 2011).  The vertical profile of the
chromospheric magnetic field can therefore be determined as \be B_z(z) =
\sqrt{8~\pi~P_{\rm gas}} .  \ee Once the magnetic field is known, the profile
of the Alfv\'en velocity, $V_A$, can be calculated throughout the chromosphere
as \be V_A= \frac{B_z(z)}{\sqrt{4~\pi~\rho(z)}} , \ee where $\rho(z)$ is the mass density.

Since the photospheres of giants and supergiants are convective and dense,
photospheric footpoints of longitudinal magnetic fields are forced to follow
the convective motions within the photosphere. The motions of magnetic field
lines with a frequency of the inverse turnover time of a stellar granule,
$\nu_A = H_p/V_c$, with $H_p$ as photospheric pressure scale height and $V_c$
as convective velocity, are able to excite MHD waves along or across the
magnetic flux tube, including torsional or transverse Alfv\'en waves (Ruderman
et al.~1997).  Torsional Alfv\'en waves represent linearly
incompressible azimuthal perturbations of the plasma velocity (linked to the
azimuthal perturbations of the magnetic field) that, unlike compressible waves
(such as longitudinal MHD waves), do not disturb the plasma density.  Although
\Alfven waves were theoretically predicted in 1942, it is only relatively recently that
researchers have reported the observational detection of them in the solar
chromosphere and corona (Tomczyk et al.~2007; De Pontieu et al.~2007; Jess et
al.~2009).

Alfv\'en waves launched from the stellar photosphere propagate upward into a
gravitationally stratified atmosphere and are subject to reflection from
regions of high gradients of Alfv\'en velocity if the wave frequency,
$\nu_{A}$, is less than the critical frequency, $\nu_{\rm crit} = dV_A/dz$
(An et al.~1990).  The interaction of downward
reflected Alfv\'en waves with upward propagated ones can ignited a turbulent
cascade of Alfv\'en waves in the lower solar atmosphere and provide a viable
source for the solar coronal heating and stellar wind heating in the open field
regions (Cranmer 2011).

Reflection of Alfv\'en waves can play an important role in driving slow and
 massive winds from the Sun, giants and supergiants (An et al.~1990; Airapetian et
 al.~1998, 2000, 2010; Suzuki 2007; Cranmer 2011; Matsumoto \& Suzuki 2014).  The radial profile of the
 critical frequency therefore provides important information about the role of
 the heating and momentum deposition of Alfv\'en waves in the atmosphere.  The
 critical Alfv\'en frequency can be calculated directly from a semi-empirical
 model by differentiating the Alfv\'en velocity profile given by
 Equation~(2).  We can calculate the gradient of \Alfven speed in the stellar chromosphere
 of a K5 giant ($\alpha$ Tau) using the semi-empirical atmospheric model of
 McMurry (1999) threaded by an open longitudinal magnetic field ($B_z$ = 100 G).
 Figure 1 shows that the Alfv\'en velocity gradient reaches its
 maximum at 0.21 $R_{\star}$.  The figure also suggests that waves at
 frequencies less than 2 mHz are trapped in the chromosphere of $\alpha$~Tau within the extent of the chromosphere characterized by
 the McMurry's model. The plot also suggests that waves at frequencies lower than 0.5$\mu$Hz are trapped in the chromosphere at heights below
 0.1 R$_{\star}$.

\begin{figure}[h]
\includegraphics[width=\linewidth]{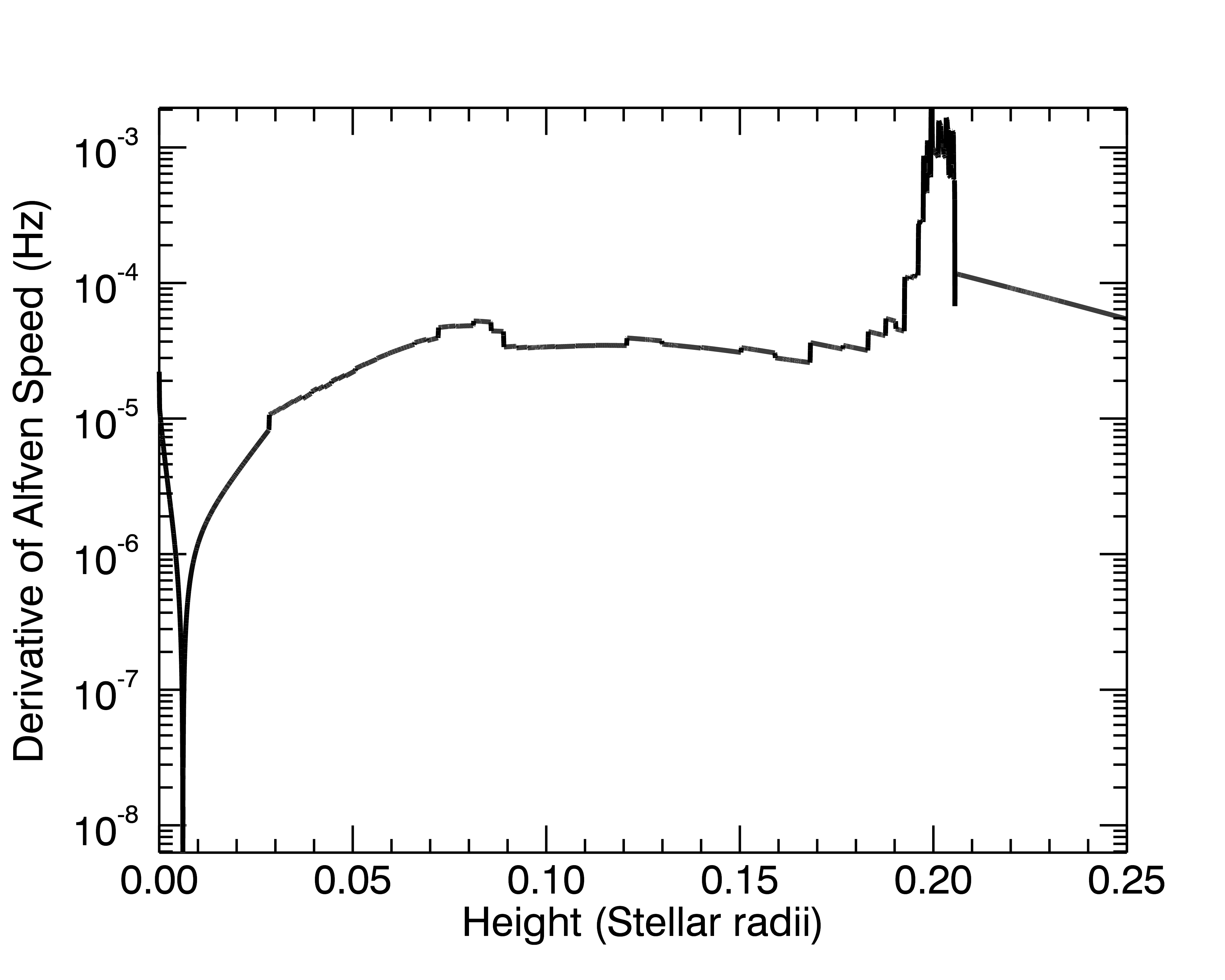}
\caption{Radial profile of the critical frequency in a stellar
chromosphere as predicted by the semi-empirical model of $\alpha$ Tau}
\end{figure}

Figure 1 suggests that the characteristic frequency of \Alfven waves launched from the photosphere
of $\alpha$ Tau should be higher than 0.5$\mu$Hz.

The magnetic field and the Alfv\'en velocity profile can also be probed by
using the Poynting theorem (Jackson 1999): \be \frac{\partial{W}}{\partial{t}}
+ \vec{\nabla} \vec{S} = - \vec{E} \vec{j} , \ee where $W =
\frac{1}{8\pi}(E^2+B^2)$ is the electromagnetic energy density and
$\vec{S}=\frac{1}{4\pi}~\vec{E}\times \vec{B}$ is the Poynting vector of the
energy source.  $\vec{S}$ represents the Poynting flux of Alfv\'en waves
launched from the photosphere.  For a steady-state chromosphere,
$\frac{\partial{W}}{\partial{t}}$ = 0 and $\vec{S}$ has only an upward
component $S_z$.  We thus obtain \be \frac{dS_z}{dz} = - <q> ,
\ee where $<q>$ is the time-averaged heating rate at a given height, $z$ (see
also Song \& Vasyliunas 2011).

The heating rate of the plasma can be derived from the energy equation for a
steady-state chromosphere where the heating rate is balanced by the thermal
conductive and radiative cooling rates, referred to as $L_{\rm cond}$ and
$L_{\rm rad}$, respectively, or

\be
- <q> = L_{\rm cond} + L_{\rm rad} .
\ee

In a stellar chromosphere, $T<0.5$~MK; the thermal conduction
time is therefore much longer than the radiative cooling time, and the thermal
conduction cooling term can be safely neglected.  Consequently, the radial
profile of the observationally derived cooling rates provides direct clues
about the profile of the Poynting flux of the heating energy source.  Detailed
information about the radial profiles of the chromospheric magnetic field and
the Alfv\'en velocity can be obtained if it is assumed that Alfv\'en waves are
the major source of the chromospheric heating.  The observations of
non-thermally broadened chromospheric lines also imply that Alfv\'en waves may
be the dominant source of energy and wind acceleration in cool giants and
supergiants; further relevant discussions have been given by Airapetian et
al.~(2010) and by Cranmer \& Saar (2011).  This type of incompressible
transverse wave can be directly excited, presumably through the shuffling or
twisting of magnetic flux tubes by well developed magneto-convection in stellar
photospheres (Ruderman et al.~1997).

The energy flux of Alfv\'en waves excited at the photosphere is defined by the
$z$-component of the Poynting vector $\vec{S} = \frac{1}{4\pi}~
\vec{E}~\times~\vec{B}$.  By applying Ohm's law, $\vec{E} = \eta \vec{j} -
\vec{V}\times\vec{B}$, Ampere's law, $\vec{j} = \frac {1}{4\pi} \vec{\nabla}
\times \vec{B}$, and using vector identities, we can write the upward Poynting
flux in Alfv\'en waves as

\be
\vec{S} = \frac{1}{4~\pi} [\vec{V}~B^2 - \vec{B}(\vec{V} \cdot\ \vec{B})] +
\frac{\eta}{4\pi}(\vec{\nabla} \times \vec{B}) \times {\vec{B} .}
\ee

If we further assume the existence of the azimuthal component only of the
velocity of footpoint motions, $V_{\phi} \neq 0$, i.e., that there are no
vertical motions in the photosphere (so $V_z = 0$), and if we represent the
total magnetic field as the sum of the background longitudinal flux-tube
magnetic field $B_z$ plus the perturbed field $\delta{B}$ due to Alfv\'en
waves, we obtain the $z$-component of the upward Poynting flux as

\be
S_z = -\frac{1}{4~\pi}~B_z~V_{\phi}~{\delta B} - \frac{\eta}{4~\pi}
{\delta B} \frac{\partial{\delta B}}{\partial{z}} .
\ee

For high magnetic Reynolds numbers, ${\rm Re}_m = \frac{V_A L}{\eta}$ ($\eta$
is the magnetic diffusivity), in the stellar chromosphere ($>$ 10, the second term in
Equation~(7) can be neglected with
respect to the first term.  Then, following the Walen relation
$\delta V_A = \delta \frac{B}{\sqrt{4\pi\rho}}$ and assuming that waves are
incompressible (so $\delta \rho$ = 0), we obtain

\be
\frac {\delta {V}}{V_A} = \frac{\delta{B}}{B_z} .
\ee

This assumption is valid until Alfv\'en waves become strongly non-linear and
convert a significant fraction of their energy into longitudinal waves (Ofman
\& Davila 1997; Suzuki 2013; Airapetian et al.~2014).  Substituting $\delta B$
from Equation~(8) into Equation~(7), we obtain the
Poynting flux as

\be
S_z = \rho~<\delta V^2> V_A .
\ee
Furthermore, when combining Eqs.(4), (5) and (9), we obtain the following:

\be
\frac{d}{dz}(\rho~<\delta V^2> V_A) = - L_{\rm rad}(z),
\ee

Equation (10) relates the thermodynamic quantities such as the plasma density,
turbulent velocity and the radiative cooling rates, which are obtained from
semi-empirical models, to the hitherto unknown vertical profile of the Alfv\'en
velocity.  Equation 10 can be rewritten as

\begin{equation}
e_A~\frac{dV_A}{dz} + \frac{de_A}{dz}~V_A = - L_{\rm rad},
\end{equation}

\noindent
where  $e_A=\rho~<\delta V^2>$ is the energy density of Alfv\'en wave energy.

Once $V_A$ is known, the profile of the magnetic field throughout the
chromosphere can be determined. {\it Hence, the knowledge of $V_A$ and
subsequent retrieval of $B_z(z)$ represents the missing link between
thermodynamic-based semi-empirical models and MHD-based theoretical models of
chromospheres and winds.} This last equation allows us to determine the range
of critical frequencies at which Alfv\'en waves become reflected from regions
where the Alfv\'en velocity gradient is at a maximum.

Comparing the magnetic-field profiles derived from Equation (11) with the one
obtained from Equation~(1) enables us to determine the degree of deviation of
the magnetic field in a chromosphere from the longitudinal (untwisted) magnetic
field, thus allowing us to constrain the value of the azimuthal magnetic field.
The magnetic-field profile in the chromosphere of $\alpha$ Tau
decreases with height at the rate of a super-radial expansion factor, $f(r)$.
Then, the magnetic field varies with the height, $r$, as $B(r)\sim
f(r)/r^2$, which is less steep than the profile obtained by Kopp \& Holzer
(1976) for solar coronal holes.

The next generation of semi-empirical models of evolved stars should therefore
combine high-resolution spectroscopic and spatial information.  Eclipsing
binaries offer a unique opportunity to derive geometric constraints on the
observed chromospheres and their winds (Eaton et al.~2008). Another promising
approach utilizes high spatial-resolution interferometric observations of
various giant and supergiant stars.

\section{1.5D MHD model of \Alfven Wave Driven Chromospheric Heating in Late Type Giants}
\subsection{Model Setup}

According to semi-empirical models of stellar chromospheres of giants (for example, McMurry 1999), the stellar chromosphere of a cool evolved star is a partially ionized environment represented by the radial profiles of the plasma temperature, $T_e$=$T_i$=$T$, the neutral density, $N_H$ and the electron density, $N_e$ (both in cm$^{-3}$) (see Left panel of Figure 2). The right panel of Figure 2 shows the radial profile of the neutral fraction, the ratio of neutral to total plasma density. One can see that throughout the chromosphere of $\alpha$ Tau, plasma remains weakly ionized.

In regions of a partially ionized and magnetized atmosphere, where the electron, $M_e$, and ion magnetization, $M_i$  (the ratio of electron/ion cyclotron frequency to the total collision frequency of electrons/ions with neutrals), both become $>$ 1 (see left panel of Figure 3), the plasma resistivity becomes anisotropic due to ambipolar diffusion via ion-neutral coupling (Michner \& Kruger 1973). First, the Spitzer resistivity, which is parallel to the magnetic field, should be modified from the fully ionized value by electron-neutral collisions. Second, the perpendicular component of the anisotropic electrical resistivity tensor, the Pedersen resistivity, $\eta_{per} \sim M_e M_i$ times the neutral fraction (right panel of Figure 3) becomes significant. This can be calculated from a NLTE Saha equation.

\begin{figure}[h]
\includegraphics*[width=0.49\linewidth]{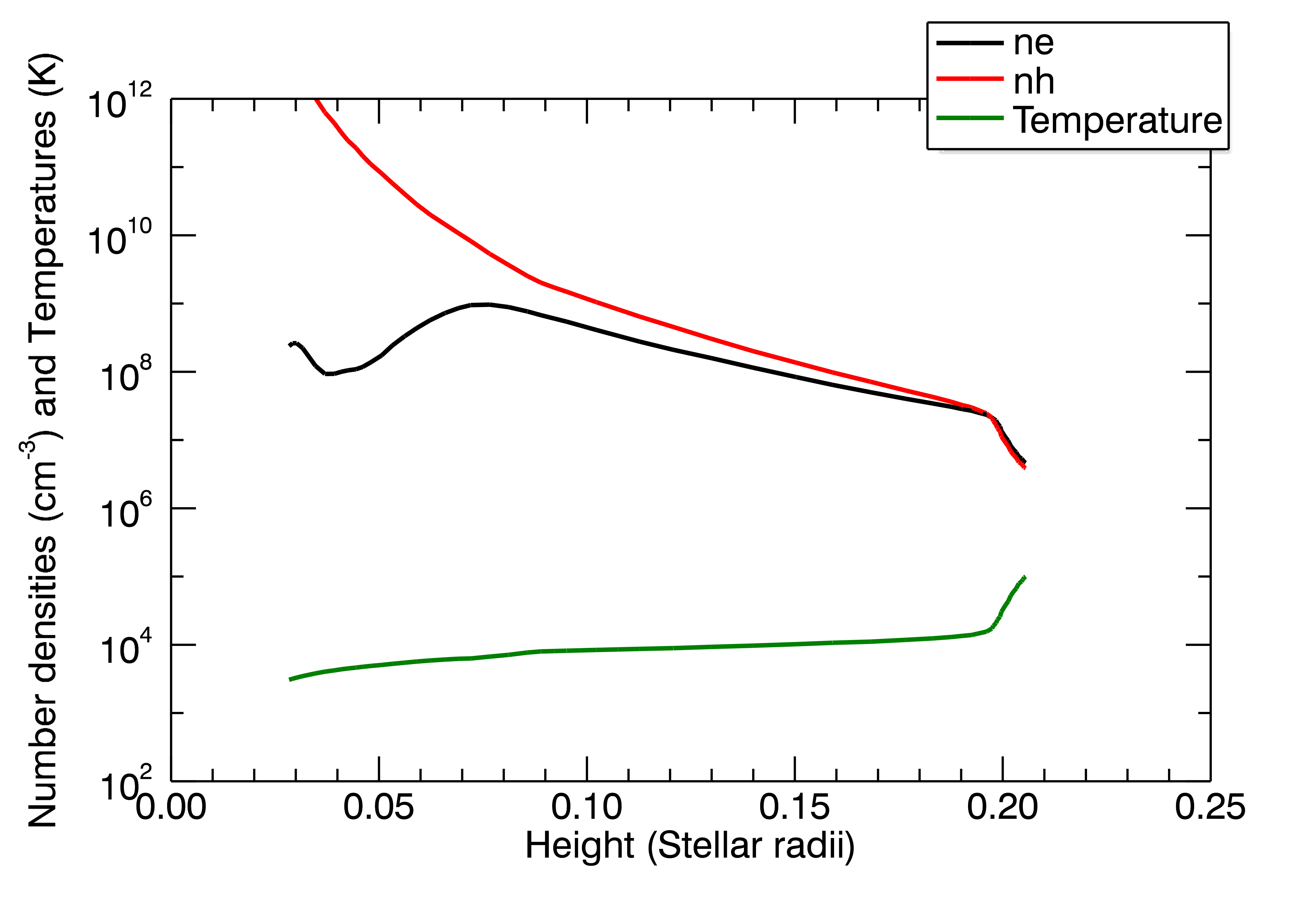}
\includegraphics*[width=0.46\linewidth]{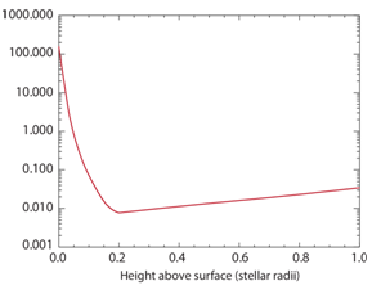}
\caption{Chromospheric model of $\alpha$ Tau. Left panel: The radial profiles of $T$, $N_H$, $N_e$; Right panel: Radial profile of neutral fraction}
\end{figure}

The left panel of Figure 3 shows that the electron and ion magnetization throughout the stellar chromosphere are much larger than 1. The left panel of the figure shows that the Pedersen resistivity is up to 8 orders of magnitudes greater than the Spitzer resistivity in the chromosphere of a giant star. It is important to note that at the grid resolution of 1500 km applied in our simulations, the numerical resistivity (green curve in the right panel of Figure 3) is 1-2 orders of magnitude smaller than the physical (Pedersen) resistivity (red curve). In most of multidimensional simulations numerical dissipation exceeds physical dissipation, so that the resistive heating rates cannot be computed accurately.  These types of simulations smooth out velocity and magnetic field gradients, and, therefore, damp electric currents.  \textbf{ Thus, our model fully resolves Pedersen resistivity and, therefore, calculates physically meaningful Joule heating rates including the contribution due to ion and electron collisions with neutrals.}

Here we apply our 1.5D MHD code to simulate MHD dynamics of a stellar chromosphere driven by upward propagating \Alfven waves at a single frequency, 0.1 \emph{mH}, launched from the photosphere with the amplitude $\delta V$ = 0.5 km/s and the surface magnetic field of 100 G along vertically diverging flux tube.

The single fluid fully non-linear resistive and viscous MHD equations in non-relativistic partially-ionized plasma are as follows:
\begin{eqnarray}
\frac{\partial \, \rho }{\partial \, t} +\nabla \cdot \left(\rho
\, \vec{V}\right)& =& 0, \label{cont:eq}\\
\rho \left[\frac{\partial\, \vec{V}}{\partial \, t} +\left(\vec{V}\cdot \nabla \right)\,
\vec{V}\right]& =& -\nabla \, p +\vec{J}\times \vec{B}+ \nabla \vec{S} \label{mom:eq} \\
\frac{\partial \, \vec{B}}{\partial \, t} & =& -\nabla \times \vec{E},\\
 \vec{E}& =& - \vec{V}\times \vec{B} + \left( \eta_{par} \vec{J}_{par} + \eta_{per} \vec{J}_{per}, \right) \label{E:eq} \\
\frac{\partial \, (\rho E)}{\partial \, t}+ \nabla (\rho E \vec{V}) & = & - P \nabla {\vec{V}} + \left( \eta_{par}~{J}_{\|}^2 + \eta_{per}~{J}_{\bot}^2 + \zeta_{ij} S_{ij} \right)
\end{eqnarray}

\noindent
Here $S_{ij}$ are the components of the stress tensor $\vec S = \nu [\zeta_{ij}-(\delta_{ij} \nabla {\vec V})/3]$ and
$\zeta_{ij}=\frac{1}{2} (\frac{\partial \,  V_i}{\partial \, x_j} + \frac{\partial \,  V_j}{\partial \, x_i})$; $\nu = \nu_{nn} + \nu_{in}$ is the viscosity coefficient due to neutral-neutral, ion-ion and ion-neutral collisions, E is the specific internal energy, $E = \frac{P}{\rho(\gamma-1)}+(1-\xi_n) \frac {X_i}{m_{av}}$. The method of solving continuity, momentum and induction equations in a partially ionized plasma has been described in detail by Arber et al. (2001) and Leake and Arber (2006).

\begin{figure}[h!]
\includegraphics*[width=0.49\linewidth]{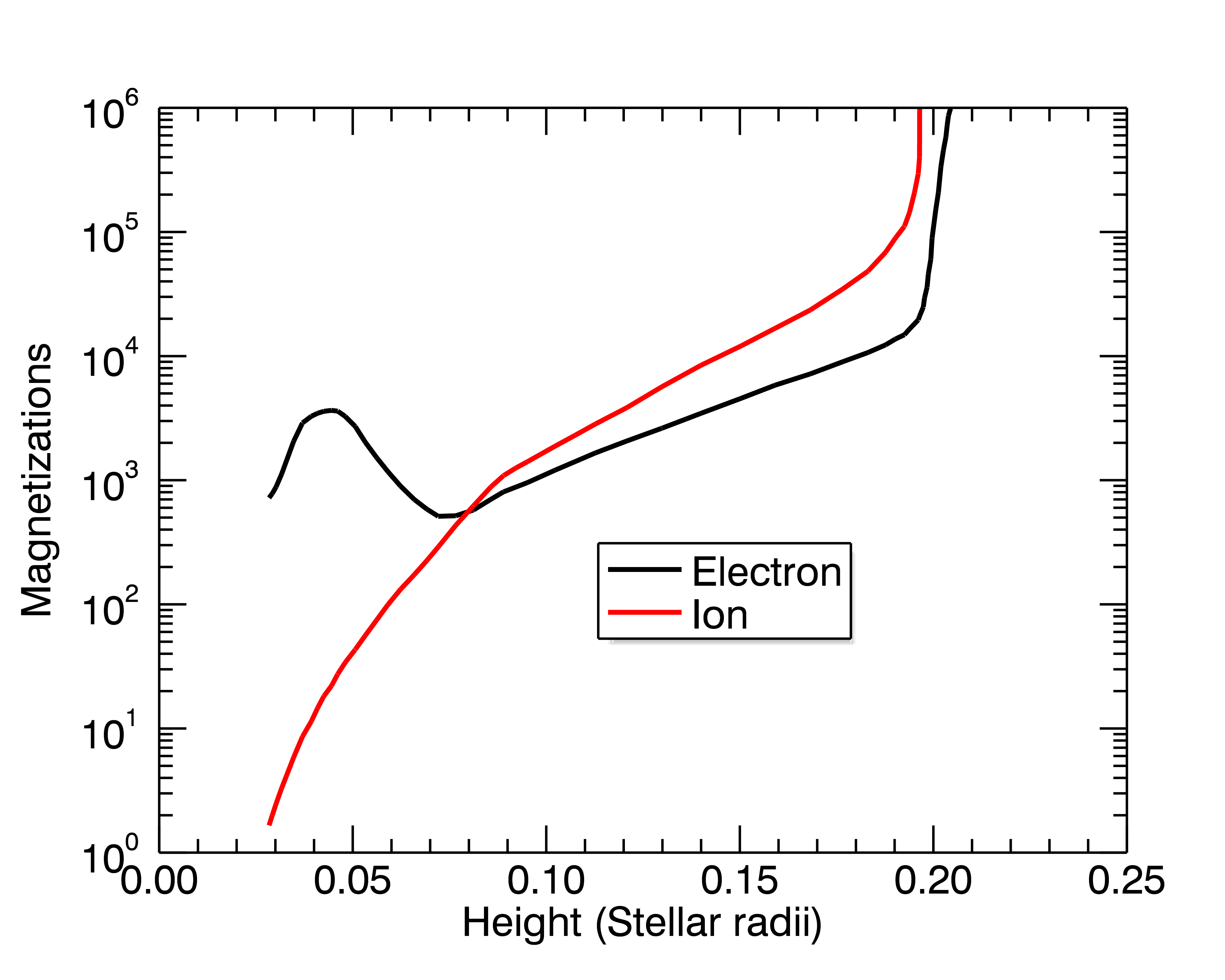}
\includegraphics*[width=0.49\linewidth]{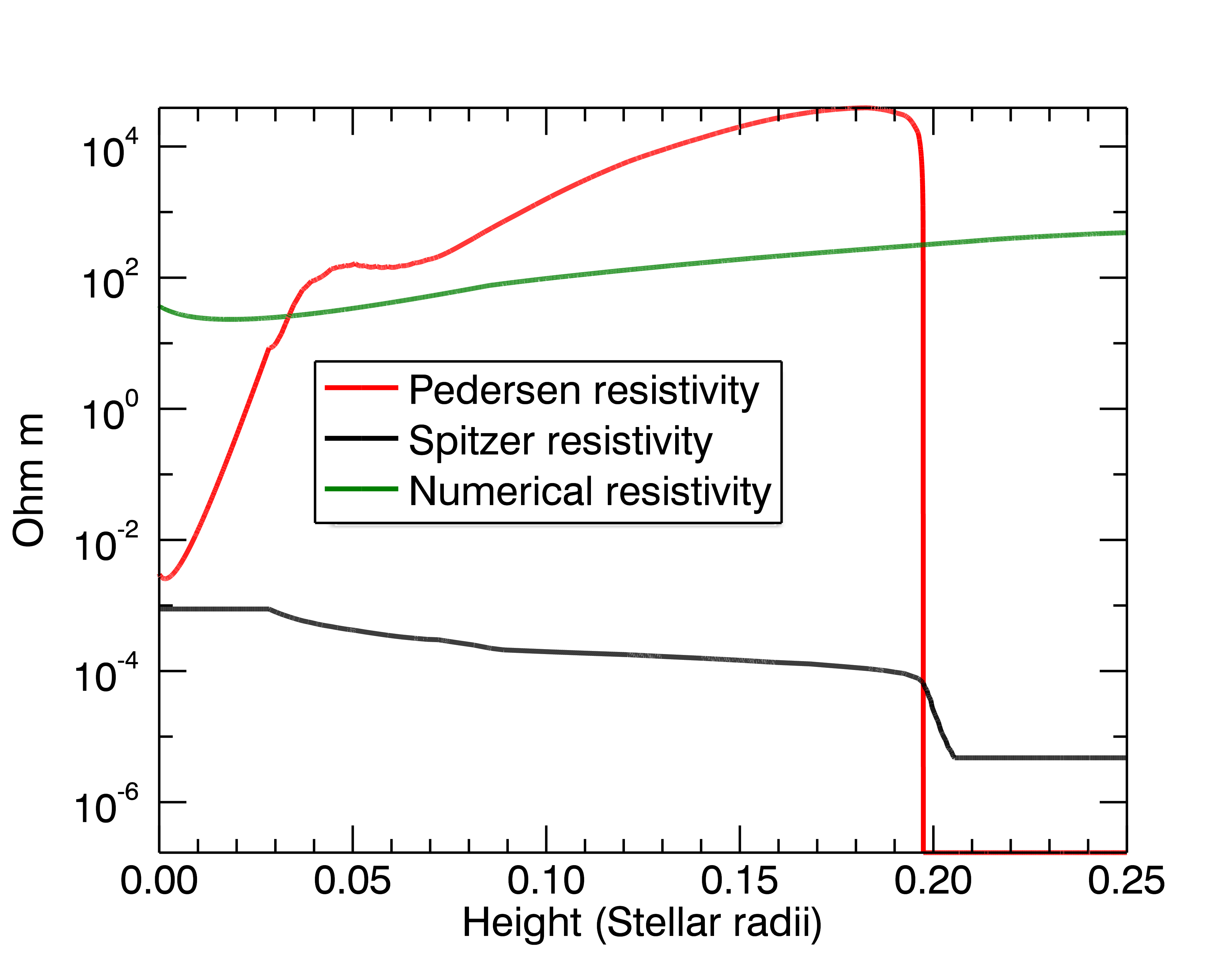}
\caption{Chromospheric model of $\alpha$ Tau. Left panel: The radial profiles of electron and ion magnetization; Right panel: Spitzer (black), Pedersen (red) and numerical resistivity (green) throughout the stellar chromosphere of $\alpha$ Tau.}
\end{figure}

\subsection{Simulation results: Energy Dissipation and Momentum Deposition in the Stellar Chromosphere of $\alpha$ Tau}

As convectively excited linear \Alfven waves in the photosphere propagate upward along the magnetic field lines, the linear theory predicts that the wave amplitude grows as density drops with height. Left panel of Figure 4 shows that the amplitude of upward propagating \Alfven waves increases by a factor 40 reaching $\sim$ 20 km/s at t=0.2 t$_A$ (green curve), where t$_A$ = 1.19$\times$10$^7$ s is the \Alfven transit time. Such wave fluctuations with $\delta V$ = 40 km/s can be a source of non-thermal turbulence in the stellar chromosphere implied from non-thermal broadening and enhanced wings (compared to a single Gaussian) observed in a number of chromospheric lines in red giant and supergiant stars. For example, the observed full width half maximum (FWHM) of optically thin C II] UV multiplet emission lines forming in the chromosphere of $\alpha$ Tau is 24 $\pm$ 1 km/s, which is much greater than that required for thermal broadening of these lines (Robinson et al. 1998) and the photospheric turbulence. The authors suggested that the enhanced wings observed in these UV lines can be explained by anisotropic turbulence directed preferentially either perpendicular or along the radial (the line of sight) direction. For the turbulence directed perpendicular to the radial direction, the limb brightening will enhance the wings of the observed profile and output the non-thermal turbulent velocity of 21 km/s, while for the radially directed turbulence, the core of the spectral line is enhanced resulting in greater turbulent velocity of 28 km/s. If we assume that the contribution in the non-thermal part of the FWHM comes from the turbulence due to unresolved \Alfven wave motions in the stellar chromosphere, then the non-thermal turbulent velocity can be directly related to the root mean square (rms) of the velocity perturbations (averaged over time greater than the wave period) caused by \Alfven waves propagating along the open magnetic field as
\be
V_{turb} = \frac{1}{2} <\delta V^2>^{1/2}~|cos{\alpha}|,
\ee

\noindent
where $\alpha$ is the angle between the plane of the transverse wave motions (perpendicular to the magnetic field) and the line of sight. Thus, for the
radially directed turbulence, $V_{turb} = 0.5{<\delta V^2>}^{1/2}$. Our model outputs the $\delta V \sim$ 40 km/s in the chromosphere, which is consistent with the observationally derived turbulent broadening in UV lines of $\sim$ 20 km/s.

At heights greater than 0.1 R$_{\star}$, the amplitude of \Alfven wave induced motions becomes comparable to the \Alfven wave speed and, therefore, the \Alfven wave motions become strongly non-linear. Such large amplitudes of transverse wave motions along the background magnetic field, $B_z$, introduce significant convective electric field, $\vec E \sim \vec \delta V \times \vec B$. The induced perpendicular component of the electric current (with respect to the vertical magnetic field) is then efficiently dissipated by the Pedersen resistivity with the volumetric Joule heating rate, $\eta_{per} J_{per}^2$. \\

\begin{figure}[h!]
\includegraphics*[width=\linewidth]{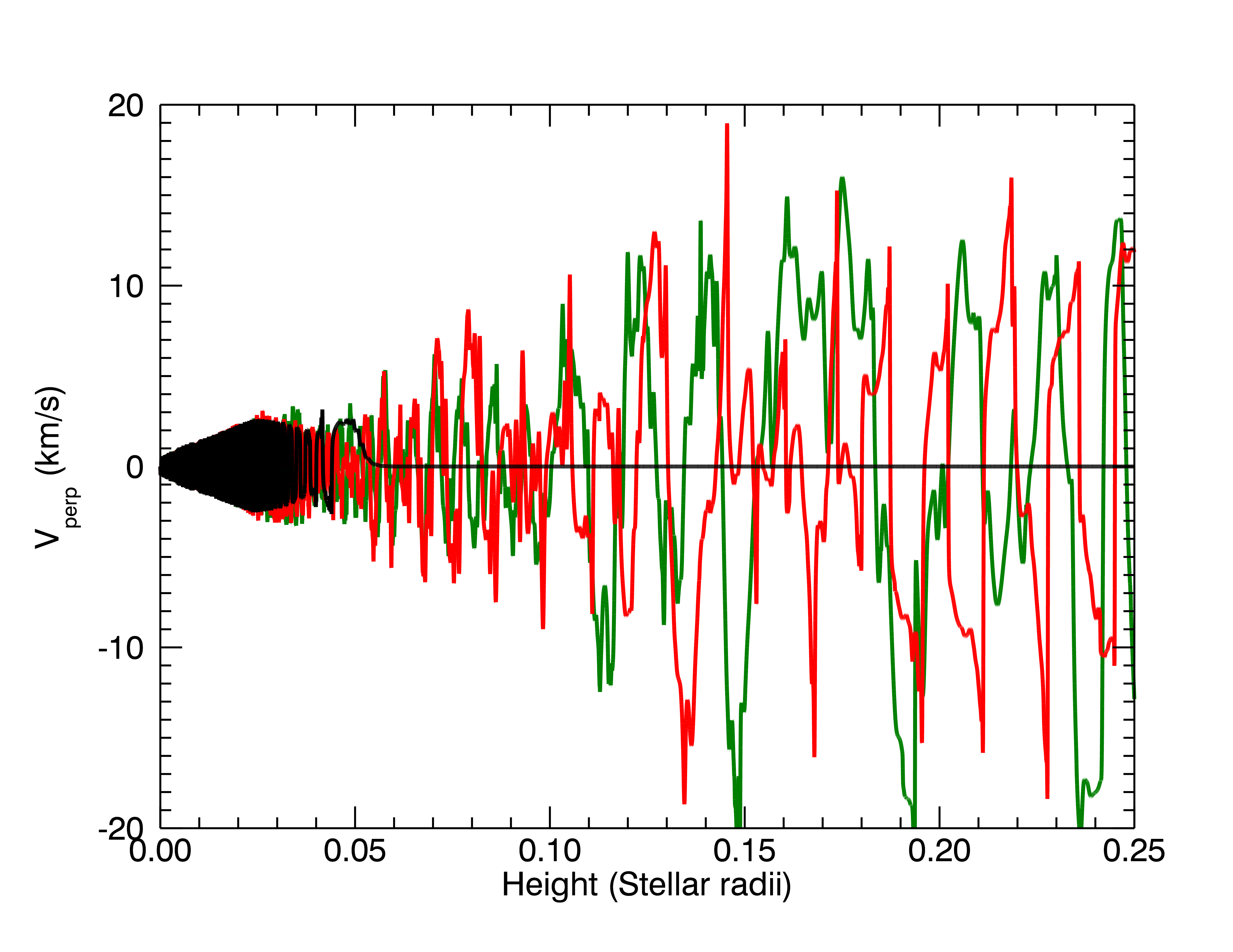}
\caption{Model outputs. Left panel: Radial profile of the wave amplitude at 0.1$t_A$ (black), 0.15$t_A$ (red) and 0.2 $t_A$ (green)}
\end{figure}

The left panel of Figure 5 shows the wave driven volumetric Joule heating rate at  0.1$t_A$ (black), 0.15$t_A$ (red) and 0.2 $t_A$ (green). The steady-state heating rate is $\sim$ 10$^{-8}$ erg/cm$^2$/s throughout the chromosphere. The right panel of Figure 5 presents the radial profile of the height integrated heating rate that reaches the peak value of 2$\times$10$^6$ ergs/cm$^2$/s at h=0.003 R$_{\star}$ from the photosphere and remains flat above this height. The heating rate is equal to the radiative cooling rate in a steady state chromosphere. The total radiative flux in the wavelength range between 1300\AA~and 3000\AA~(continuum and in the Mg II h\&k lines) from $\alpha$ Tau is about $\sim$10$^6$ ergs/cm$^2$/s (Robinson et al. 1998). Thus, \Alfven wave dissipation in our model provides enough heating flux to balance the radiative losses at the filling factor of "magnetic active regions" $\leq$ 1!

\begin{figure}[h!]
\includegraphics*[width=0.49\linewidth]{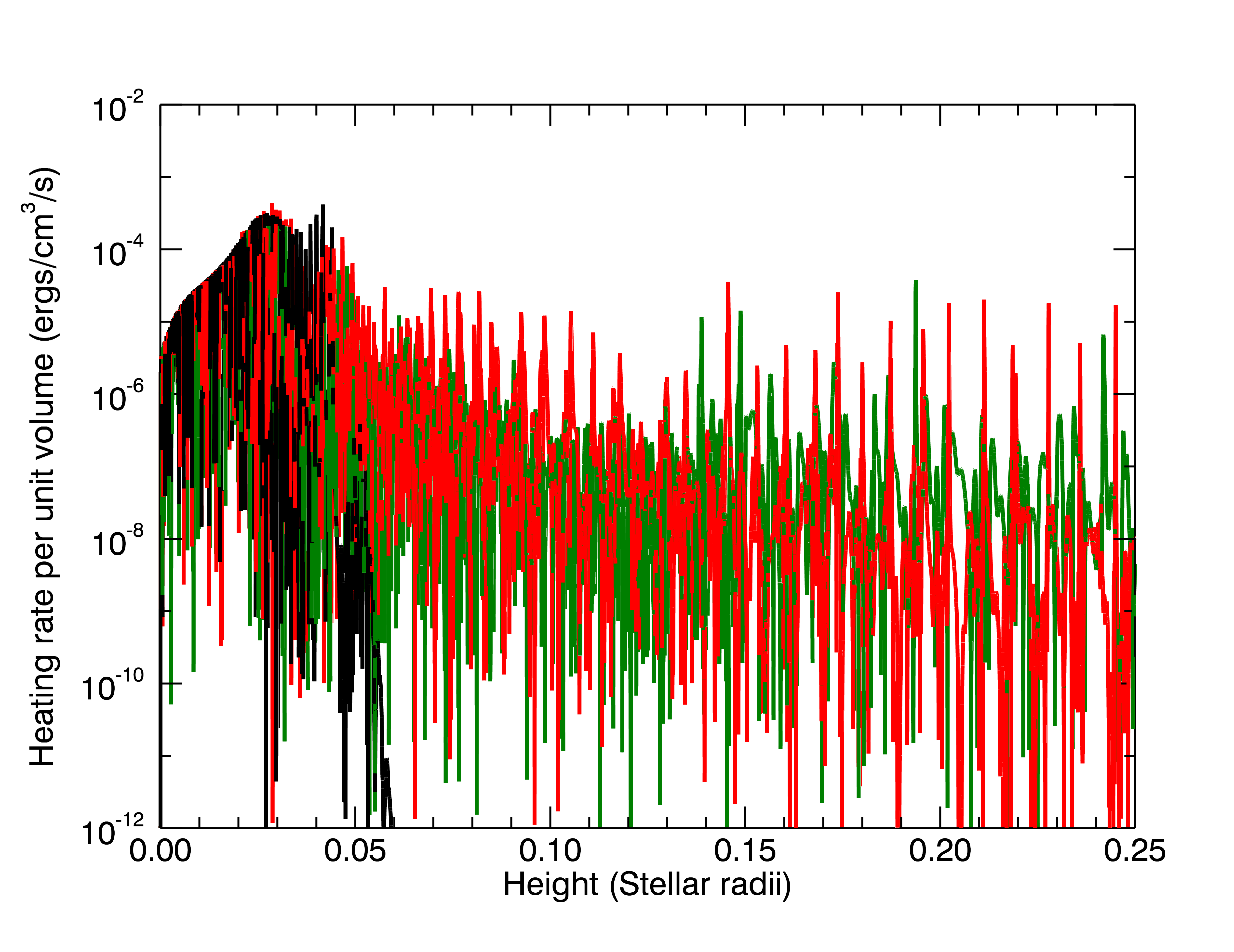}
\includegraphics*[width=0.49\linewidth]{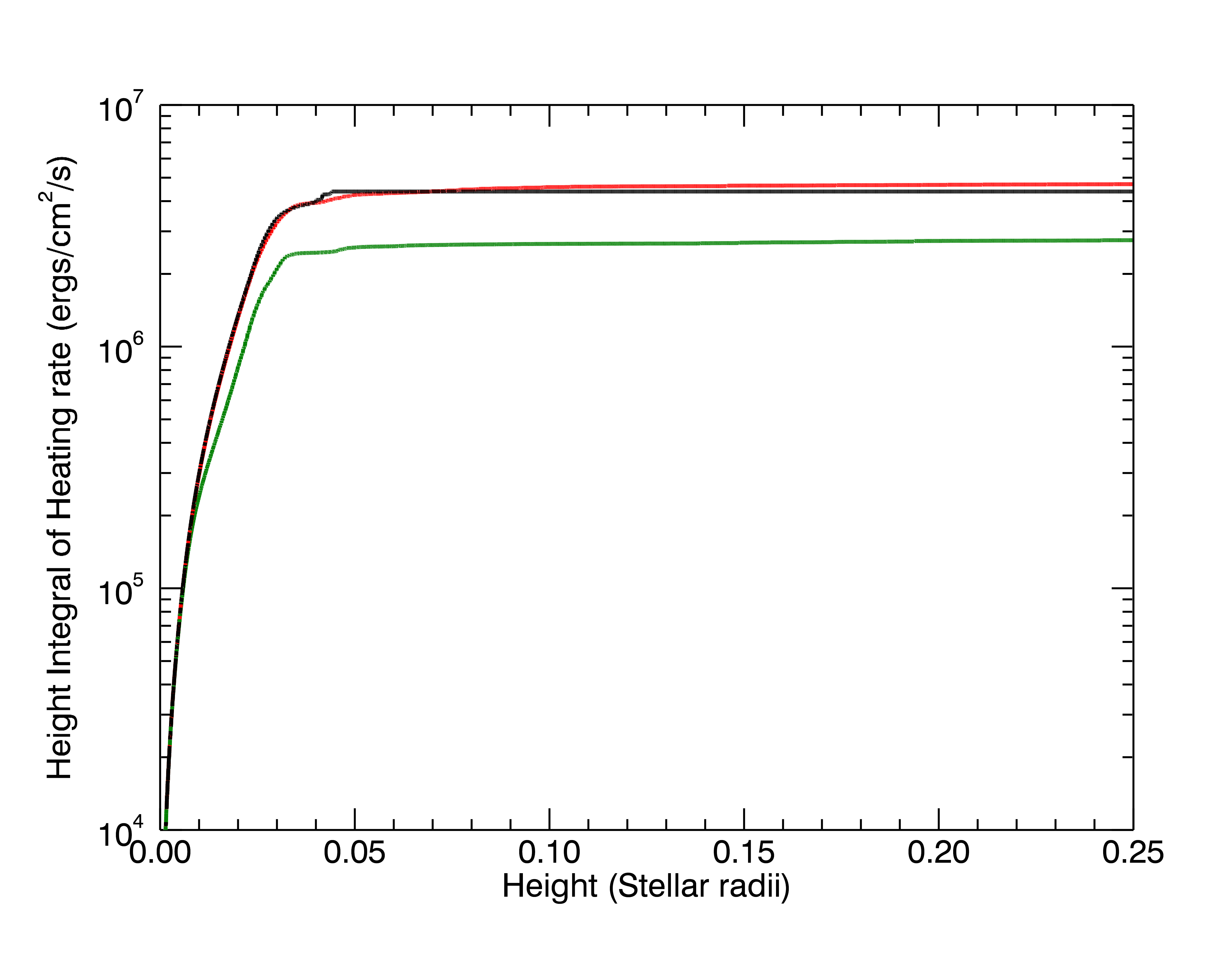}
\caption{Model outputs. Left panel: Radial profile of the volumetric (left panel) and integrated over the atmospheric height (right panel) Joule heating at 0.1$t_A$ (black), 0.15$t_A$ (red) and 0.2$t_A$ (green)}
\end{figure}

The deposition of the momentum of the \Alfven waves occurs through the wave-generated Lorentz force, $\vec{J}\times\vec{B}$, which is exerted on the plasma (expressed by the momentum equation.
This wave-generated force provides plasma acceleration through the gradient of
the \Alfven wave pressure, $\frac{1}{\rho}~\nabla(\frac{B^2}{8\pi})$, and by the magnetic tension force,
$\frac{1}{4\pi~\rho}~(\vec{B}\cdot\vec{\nabla}){\vec{B}}$. The heating of the wave creates the gradient of the plasma pressure, $\nabla P$, in the
momentum equation, and can therefore also provide additional acceleration to drive
the winds from the Sun and coronal giants. However, this term is not significant for
accelerating cool winds from non-coronal giants and supergiants, such as $\alpha$ Tau and $\alpha$ Ori.

\begin{figure}[h!]
\includegraphics*[width=0.49\linewidth]{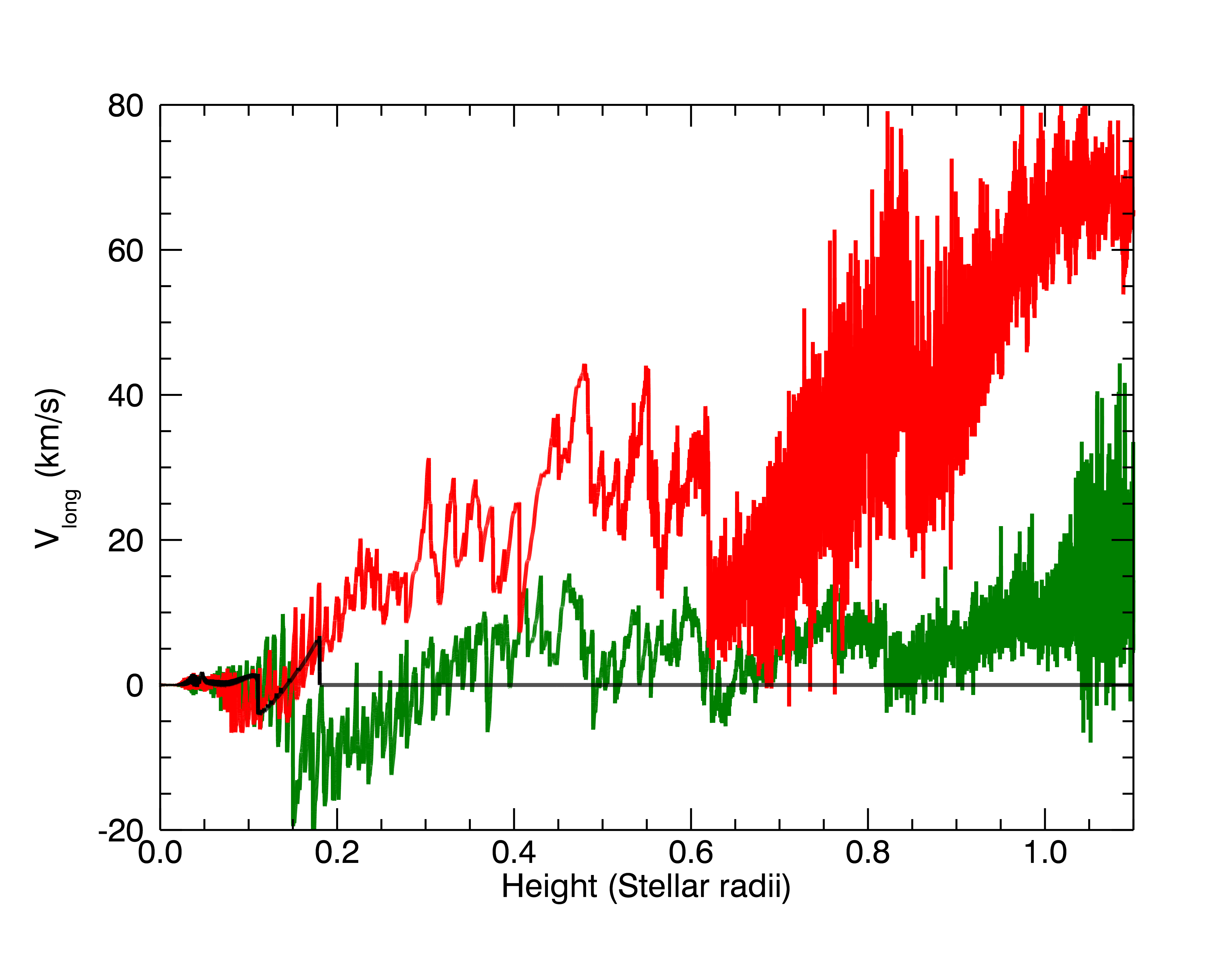}
\includegraphics*[width=0.49\linewidth]{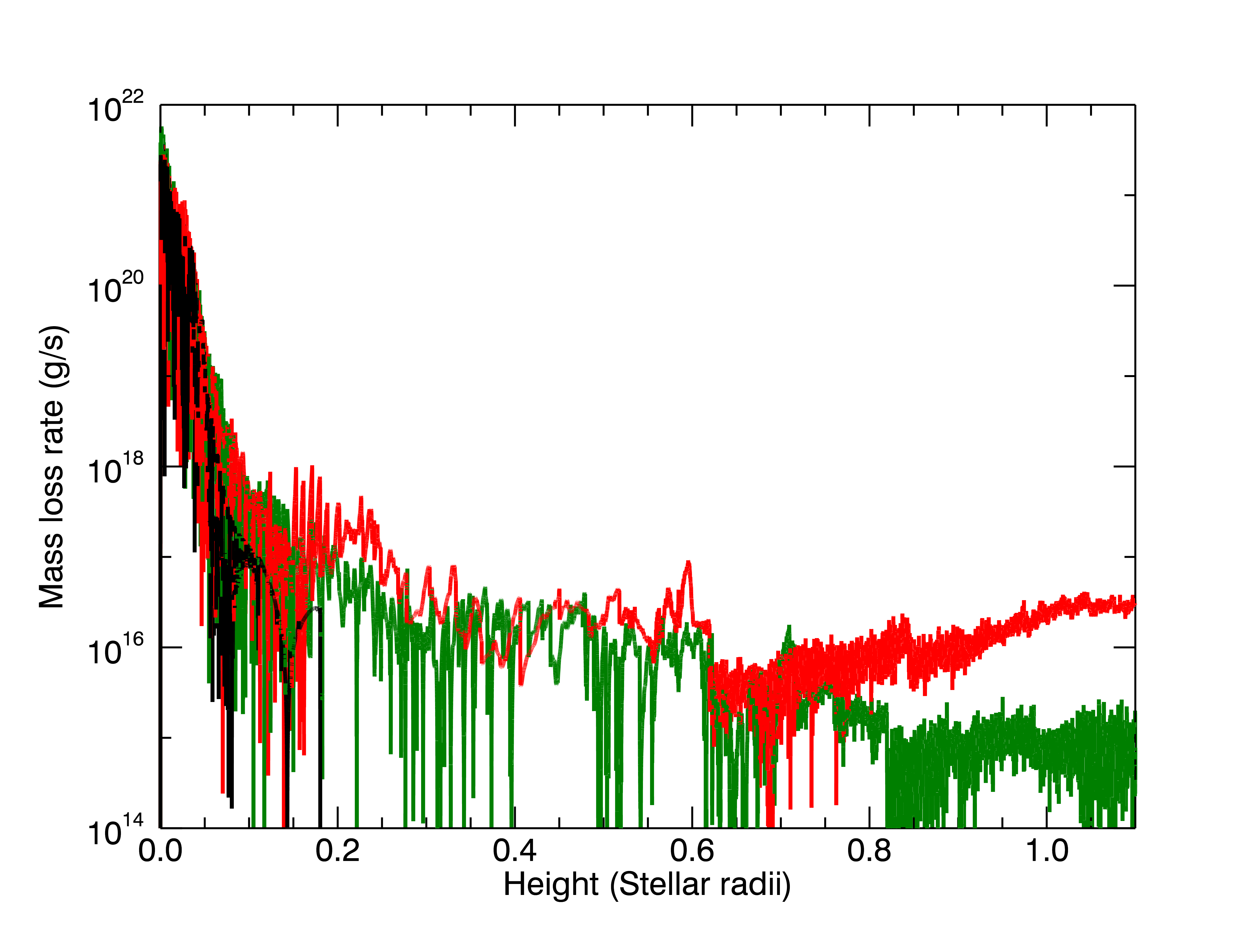}
\caption{Model outputs. Left panel: Radial profile of the radial velocity of atmospheric plasma (left panel) and mass loss rate due to \Alfven wave accelerated wind at 0.1$t_A$ (black), 0.15$t_A$ (red) and 0.2$t_A$ (green)}
\end{figure}

Figure 6 shows the radial profiles of the chromospheric radial velocity, $V_{long}$ (left panel of Figure 6) and the radial momentum deposition in the chromosphere in terms of the mass loss rate, $dM/dt = 4\pi r^2\rho V_{long}$ ($r$ is radial distance from the photosphere), throughout the atmosphere at three time moments t=0.1 (black), 0.15 (red) and 0.2 t$_{A}$ (green). Figure 6 shows that the wind starts accelerating in the chromosphere at the top of the chromosphere $\sim$0.2 $R_{\star}$ and reaches the terminal velocity of $\sim$ 30 kms/s at about 1 R$_{\star}$. The plot shows that the mass loss rate becomes flat at about 2$\times$10$^{15}$ g/s = 3.17$\times$10$^{-11}$ M$_{\odot}$/yr at heights $\sim$ 1.1 R$_{\star}$. This suggests that in order to explain the observationally derived mass loss rate, $\sim$1.6$\times$10$^{-11}$ M$_{\odot}$/yr, the filling factor of the open magnetic field should be $\leq$ 1, which is consistent with the filling factor required to explain the radiative losses as discussed earlier.

\section{Conclusions}

We have performed the first self-consistent MHD modeling with partial ionization of the chromospheric heating and wind acceleration driven by \Alfven waves launched from the stellar photosphere of a typical red giant, $\alpha$ Tau. The \Alfven wave driven energy dissipation and momentum deposition in the atmosphere of $\alpha$ Tau are consistent with observational signatures in red giant and supergiant stars. First, our model predicts the presence of non-thermal broadening and enhanced wings observed in optically thin UV lines of late-type giants and supergians, as the result of the anisotropic large-scale turbulent motions introduced by unresolved non-linear transverse \Alfven waves propagating along an open magnetic field, the source of anisotropy. Next, the model suggests that such large amplitudes of the non-linear wave motions perpendicular to the background field in a partially ionized stellar chromosphere introduce perpendicular electrical currents. These electric currents can be efficiently dissipated as Joule heating via Pedersen resistivity and explain the observed heating rates in the chromosphere of $\alpha$ Tau deduced from the net cooling losses in UV emission lines and continuum. Most MHD simulations describe dissipation processes relevant to a fully ionized plasma, where numerical dissipation greatly exceeds physical dissipation, so that the resistive rates cannot be computed accurately. Because we have included the effects of ion-neutral collisions on plasma resistivity and a fine computational grid throughout the chromosphere, we have been able to fully resolve the resistivity. Our simulations also show that the Lorentz force exerted by \Alfven waves on chromospheric plasma above the top of the chromosphere can explain the plasma acceleration to the terminal velocity, which is consistent with the observationally derived terminal wind velocity from $\alpha$ Tau (Robinson et al. 1998). In addition, the theoretically derived mass loss rate is also in quantitative agreement with the mass loss rates derived from observations of UV lines forming in the wind of $\alpha$ Tau (Carpenter and Robinson 1995;  Robinson et al. 1998).

Thus, our numerical 1.5D MHD model that launches \Alfven waves directly from the photosphere in a gravitationally stratified atmosphere of $\alpha$ Tau can consistently explain the turbulent velocities observed in non-thermally broadened UV line and relate this turbulence to non-linear \Alfven waves. These waves then dissipate enough energy to explain radiative losses in the chromosphere and deposit enough momentum to drive slow (20 km/d) and massive ($1.5\times10^{-11} M_{\odot}$/yr) winds in outer chromosphere above 1 stellar radius from the stellar surface.
Detailed examination of chromospheric emission lines of
Fe II, O I and Mg II indicate that the wind from a late-type giant including $\alpha$ Tau appears to originate near the base of the chromosphere and continues to accelerate throughout the entire chromospheric
region (Carpenter et al. 1995). It is therefore assumed that the wind reaches its terminal velocity within one stellar radius as expected from our MHD simulations.

In our next study, we will include the radiative cooling term to construct a realistic thermodynamic MHD model of the stellar chromosphere and to calculate fluxes in Ca II and Mg II emission lines that can be directly compared with observations.

\acknowledgments{
V. A. acknowledges financial support from the NASA grant NNG09EQ01C.
}

\end{document}